%% file: two_phase_shocktube.tex
\documentclass[preprint,review,12pt]{elsarticle}

\usepackage{amssymb}

\include{packages}
\include{tikz_settings}
\include{commands}

\journal{Journal of Computational Physics}

\begin{document}

\begin{frontmatter}



\title{Comparison of Macro- and Microscopic Solutions of the Riemann Problem II. Two-Phase Shock Tube}


\author[label1]{Timon Hitz}
\author[label1]{Steven J\"ons}
\ead{joens@iag.uni-stuttgart.de}
\author[label2]{Matthias Heinen}
\author[label2]{Jadran Vrabec}
\author[label1]{Claus-Dieter Munz}

\address[label1]{Institute of Aerodynamics and Gas Dynamics, University of Stuttgart, 70569 Stuttgart, Germany}
\address[label2]{Thermodynamics and Process Engineering, Technical University of Berlin, 10623 Berlin, Germany}

\begin{abstract}

The Riemann problem is one of the basic building blocks for numerical methods in computational fluid mechanics.  Nonetheless, there are still open questions and gaps in theory and modelling for
situations with complex thermodynamic behavior. In this series, we compare numerical solutions of the macroscopic flow equations with molecular dynamics simulation data. To enable molecular dynamics
for sufficiently large scales in time and space, we selected the truncated and shifted Lennard-Jones potential for which also highly accurate equations of state are available.  A comparison of a
two-phase Riemann problem is shown, which involves a liquid and a vapor phase, with an undergoing phase transition. The loss of hyperbolicity allows for the occurrence of anomalous wave structures. We
successfully compare the molecular dynamics data with two macroscopic numerical solutions obtained by either assuming local phase equilibrium or by imposing a kinetic relation and allowing for
metastable states.    

\end{abstract}


\end{frontmatter}


\section{Introduction}
\label{sec:introduction}
The Riemann problem is an initial value problem with piecewise constant initial data and one of the most important building blocks to construct numerical methods for conservation equations.  Pioneered
by Godunov \cite{Godunov1959}, its consideration within finite volume (FV) schemes gives way to the construction of numerical flux functions for various applications. FV methods, equipped with an
appropriate Riemann solver, are known to be robust and accurate in solving problems exhibiting strong gradients or discontinuities \cite{toro_riemann_2009}. Next to the construction of numerical
fluxes, the Riemann problem is also a canonical test case, providing critical insight into the solution of non-linear hyperbolic differential equations (PDE) and their numerical approximations.  We
started novel considerations for compressible fluid flow with the first part of our paper \cite{hitz_comparison_2020}. Therein, we constructed the exact solution of the Riemann problem for the Euler
equations and calculated an approximate solution with a mixed DG/FV scheme based on a Riemann solver flux. We compared both with molecular dynamics simulations.  While modeling is straightforward for
an ideal gas, the solution theory for macroscopic situations still has gaps for complex thermodynamic and hydrodynamic interactions, which occur e.g. at interfaces between phases.  To enable molecular
dynamics simulations for sufficiently large scales in time and space, we selected the truncated and shifted Lennard-Jones (LJTS) potential for the intermolecular interactions. To attain a one-to-one
correspondence, an accurate equation of state (EOS) specifically derived for this potential was employed for the solution of the flow equations.   

We started in Ref. \cite{hitz_comparison_2020} with a supercritical situation. This is a regime in which the EOS for the LJTS fluid is convex and the structure of the Riemann problem solution is
well-known. For a classical, fully convex EOS,  one can establish a unique solution providing an entropy condition, e.g. the Lax criterion or Liu's entropy criterion \cite{godlewski_numerical_2014}.
Results from three different approaches showed an excellent agreement with each other, even for a jump from a liquid-like to a gas-like state in the initial data. We also considered the expansion into
vacuum. Deviations between the molecular dynamics data and the continuum solution could only be seen while approaching the critical Knudsen number and once condensation effects arose.  In summary, the
results validated the possibility of comparing micro- and macroscopic results directly. 

We now continue with this comparison for a two-phase Riemann problem  involving  a liquid and a vapor, where phase transition may occur. From the initial data in different phases, the solution must
find a path through the two-phase region. Inside the two-phase region, the isotherm and isentrope exhibit a Maxwell loop that entails an imaginary speed of sound. Due to this non-convexity of the
EOS, the hyperbolicity of the Euler equations is lost and the PDE become of hyperbolic-elliptic type.  The loss of hyperbolicity allows for the occurrence of anomalous wave structures, such as split
waves, composite waves or non-classical shock waves. Menikoff and Plohr \cite{menikoff_riemann_1989} discussed these phenomena in great detail for the Euler equations. They argue that in case of phase
transition, the solution of the Riemann problem is non-unique since the underlying equations neglect certain physical processes, such as viscous dissipation or heat conduction. 

In the case of non-convex EOS and phase transitions, the solution is more complex and the entropy condition depends on additional physical effects. Menikoff and Plohr \cite{menikoff_riemann_1989}
restricted themselves to the assumption of being in thermodynamic phase equilibrium at the interface. This approach is known as homogeneous equilibrium method (HEM). The states in the two-phase region
are assumed to be a mixture of saturated liquid and vapor at the same temperature, pressure and Gibbs energy. The EOS then suffers a kink at the coexistence curves of the two phases, which leads
to the appearance of split and composite waves. An extension of this approach was pursued by M\"uller and Vo\ss~\cite{muller_riemann_2006} as well as Vo\ss~\cite{vos_exact_2005}, using the van der
Waals EOS and a Maxwell construction.

An alternative approach introduces a sharp interface, at which jump conditions hold and a kinetic relation determines the entropy production upon phase transition. This concept was pioneered by
Abeyaratne and Knowles \cite{abeyaratne_driving_1990,abeyaratne_kinetic_1991,abeyaratne_evolution_2006} to describe the propagation of phase boundaries in solids. When considering the kinetic
relation, the non-convexity of the EOS is maintained and the interface corresponds to a non-classical undercompressive shock wave, which is not associated with any eigenvalue of the Euler equations. 

For the fully isothermal Euler equations, kinetic relation theory was adapted to interfaces between liquid and vapor
\cite{dreyer_bubbles_2011,hantke_exact_2013,rohde_relaxation_2015,rohde_riemann_2018,truskinovsky_kinks_1993}.  An extension to the full Euler equations was discussed by Fechter
\cite{fechter_compressible_2015}, Zeiler \cite{zeiler_liquid_2016}, Fechter \etal\cite{fechter_sharp_2017,fechter_approximate_2018} and Thein \cite{thein_results_2018}. These works expanded the theory
of kinetic relations towards the non-isothermal case. However, they only took the Euler equations into account. Consequently, heat conduction was neglected in the bulk phases so that different
modifications and adaptions had to be made. 

In this paper, we consider both macroscopic approaches: the HEM and the sharp interface method which are implemented in the discontinuous Galerkin spectral element method (DGSEM) framework \flexi. The
sharp interface method is implemented in one spatial dimension based on an Arbitrary-Lagrangian-Eulerian (ALE) method \cite{etienne_geometric_2008}. A description of the ALE method in the context of
DGSEM was given by Minoli and Kopriva \cite{minoli_discontinuous_2011}. The velocity of the interface is determined by solving the two-phase Riemann problem, taking heat conduction across the
interface into account by Fourier's law. Therefore, the Riemann solver of Fechter \etal \cite{fechter_sharp_2017,fechter_approximate_2018} was extended to this case.  

The structure of the paper is as follows. In Section \ref{sec:mathematical_model}, the governing equations, the EOS and the modeling approaches for HEM and sharp interface method are introduced.
The numerical methods, both continuum and molecular dynamics, are discussed in Section \ref{sec:numericalmethods}. Results for two-phase Riemann problems with phase transition are given in
Section \ref{sec:results}, followed by a conclusion in Section \ref{sec:conclusion}.

\section{Mathematical Model}
\label{sec:mathematical_model}

In the sharp interface model, the computational domain $\Omega$ contains a pure fluid that may exist in the liquid or vapor state.  Both phases are divided by a moving interface $\Gamma(t)$, which is
assumed to be infinitesimally thin and to not carry any mass or energy.  It separates $\Omega$ into two subdomains $\Omega_{\liq}(t)$ and $\Omega_{\vap}(t)$, which contain either only liquid or only
vapor, respectively.

\subsection{Continuum Equations}

The one-dimensional motion of an inviscid, but heat conducting, two-phase fluid in the domain $\Omega$ is described by a set of balance equations in the bulk phases $\Omega_{\vap}(t)$ and
$\Omega_{\liq}(t)$ and suitable jump conditions across the interface $\Gamma(t)$. The balance equations read 
\begin{align}
   \label{eq:bulk_1}
   \rho_t + \div \rho v & = 0 , \\
   \label{eq:bulk_2}
   \l( \rho v \r)_t + \div \l( \rho v^2 + p \r)  & = 0 , \\
   \label{eq:bulk_3}
   \l( \rho e \r)_t + \div \l( \l( \rho e + p \r) v \r) + \div \q  & = 0
   ,
\end{align}
and are completed by suitable initial and boundary conditions \cite{ishii_thermo-fluid_2011}.  The unknowns are density $\rho(\x,t)>0$, velocity $\v(\x,t)$ and specific total energy $e(\x,t)$, which
is the sum of specific internal energy $\epsilon$ and specific kinetic energy $e(\x,t)=\epsilon+\half v^2$. The variables $p$ and $q$ denote pressure and  heat flux, respectively. 

The jump conditions across the interface $\Gamma(t)$ read  
\begin{align}
   \label{eq:jump_1}
   \llbracket \dot{m} \rrbracket & = 0 , \\
   \label{eq:jump_2}
   \dot{m} \llbracket v \rrbracket + \llbracket p  \rrbracket & = \surfacetension , \\
   \label{eq:jump_3}
   \dot{m} \llbracket \epsilon + \frac{p}{\rho} + \half \l( v - S \r)^2 \rrbracket + \llbracket q \rrbracket & = 0
   .
\end{align}
The jump brackets for an arbitrary quantity $\phi$ are defined as $\llbracket \phi \rrbracket  = \phi_{\vap} - \phi_{\liq}$ with $\phi_{\vap/\liq}=\lim_{h \rightarrow 0,h>0}  \phi \l( \tilde{\x} \pm h
\ninterface \r)$, where $\tilde{\x}\in \Gamma(t)$ and $\ninterface$ is the normal vector of the interface at $\tilde{\x}$ . It is defined to always point towards the vapor domain.  Note that only
planar interfaces are considered here, and surface tension effects are neglected. The mass flux across the interface is given by $\dot{m}=\rho_{\liq/\vap}  \l( \v_{\liq/\vap} - S \r)$ with the
propagation speed $S$ of the interface.

In addition to the balance equations, the bulk phases in $\Omega_{\liq}(t)$ and $\Omega_{\vap}(t)$ need to satisfy an entropy inequality in the weak sense
\begin{equation}
   \label{eq:entropycondition_bulk}
   \l( \rho s \r)_t + \div \l( \rho s \v + \frac{q}{T} \r) \geq 0,
\end{equation}
according to the second law of thermodynamics. Across the interface, the  additional jump condition
\begin{equation}
   \label{eq:entropycondition_jump}
   \dot{m} \llbracket s \rrbracket + \llbracket \frac{q}{T} \rrbracket = \eta_{\Gamma}, \quad \eta_{\Gamma} \geq 0 ,
\end{equation}
has to be fulfilled, where $\eta_{\Gamma}$ is the entropy production rate at the interface.

The heat flux was modeled by Fourier's law
\begin{equation}
\label{eq:fourier}
\q = -\lambda \gradient T
,
\end{equation}
where $\lambda>0$ is the thermal conductivity.

\subsection{Equation of State}\label{sec:EOS}

To close the balance \cref{eq:bulk_1,eq:bulk_2,eq:bulk_3} and the jump conditions \cref{eq:jump_1,eq:jump_2,eq:jump_3}, an EOS is required for the considered fluid. One usually distinguishes between
thermal EOS $p=p(\rho,T)$ and caloric EOS $\epsilon=\epsilon(\rho,T)$.  For a given density, the temperature can be calculated by inverting the caloric EOS and may be used to obtain the pressure.
However,  we employ a more general formulation in terms of a thermodynamic potential, i.e. the Helmholtz energy
\begin{equation}
\label{eq:helmholtzenergy}
\psi = \psi \l( \rho,T \r)
.
\end{equation}
EOS in the form of \cref{eq:helmholtzenergy} yield all other thermodynamic quantities, such as pressure or internal energy, via differentiation only. 
 
To get a close connection of the macroscopic and microscopic simulations, we restrict ourselves to the LJTS fluid. For this model fluid, two EOS are available: the empirical multiparameter LJTS EOS by
\citet{thol_equation_2015} and the semi-empirical PeTS EOS by \citet{heier_equation_2018}.  Both provide an expression for the reduced Helmholtz energy in the form 
\begin{equation}
\label{eq:eos_helmholtz}
\frac{\psi\l(\rho,T\r)}{\Rgasconstant T} = F (\delta,\theta) = F^0 (\delta,\theta) + F^{{r}} (\delta,\theta),
\end{equation}
as a function of the reduced density $\delta=\rho/\rho_{\mathrm{c}}$ and inverse reduced temperature $\theta = T_{\mathrm{c}}/T$.  The symbol $\Rgasconstant$ denotes the gas constant.  The reduced
Helmholtz energy is composed of an ideal gas contribution, indicated by superscript "$^0$", and a residual contribution that accounts for the intermolecular interactions, indicated by the superscript
"$^{{r}}$".  This model is an adequate representation of the noble gases and methane in the entire range of fluid states \cite{Rutkai2017}. All time-independent thermodynamic quantities can be
calculated by analytical differentiation of \cref{eq:eos_helmholtz}
\begin{align}
\label{eq:helmholtz_pressure}
\text{pressure: } \quad
p & = \rho T R \l( 1+\delta F^{{r}}_{\delta} \r), \\
\label{eq:helmholtz_internalenergy}
\text{internal energy: } \quad
\epsilon & = T R \l( \theta \l( F^0_{\theta} + F^{{r}}_{\theta} \r) \r), \\
\label{eq:helmholtz_entropy}
\text{entropy: } \quad
s & = R \l( \theta \l( F^0_{\theta} + F^{{r}}_{\theta} \r) - \l( F^0 + F^{{r}} \r) \r), \\
\label{eq:helmholtz_enthalpy}
\text{enthalpy: } \quad
h & = \epsilon + p \tau = T R \l( \l( 1 + \delta F^{{r}}_{\delta} \r) + \tau \l( F^0_{\tau} + F^{{r}}_{\tau} \r) \r), \\
\label{eq:helmholtz_gibbs}
\text{Gibbs energy: } \quad
g & = h - Ts = T R \l( \l( 1 + \delta F^{{r}}_{\delta} \r) + \l( F^0 + F^{{r}} \r) \r),
\end{align}
with $F_{\delta}=\partial F/\partial \delta$ and $F_{\theta}=\partial F/\partial \theta$.

\begin{figure}[!ht]
   \centering
   \includegraphics[width=0.6\linewidth]{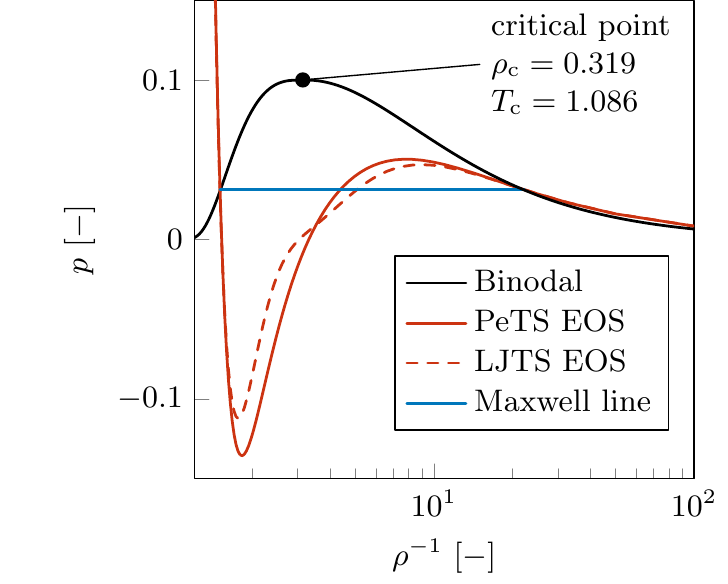}
   \caption{Pressure of the LJTS model fluid along the isotherm $T=0.9$ calculated with the PeTS and LJTS EOS, respectively.}
  \label{fig:ljts_isothermal}
\end{figure}

\Cref{fig:ljts_isothermal} shows the pressure of the LJTS fluid along the isotherm $T=0.9$ calculated with the PeTS and LJTS EOS, respectively.
Both EOS exhibit so-called Maxwell loops in the two-phase regime. They occur due to the non-convexity of the Helmholtz energy and allow for a continuous connection of the liquid phase with the vapor
phase. The PeTS EOS exhibits a single loop in the entire temperature range so that it qualitatively complies with classical cubic EOS, such as the one by Van der Waals. The two-phase region is bounded
by the binodals.  Between the spinodals, which are the loci of the extrema, imaginary values for the speed of sound occur. Between the binodals and spinodals, the states are metastable and have a
real value for the speed of sound. For the LJTS EOS, several Maxwell loops may appear. In Ref. \cite{hitz_comparison_2020}, we validated both EOS against molecular dynamics simulation data on
supercritical shock tube scenarios as well as expansion waves into vacuum.  It was found that the PeTS EOS, although being less accurate than the LJTS EOS, is a well-suited candidate to perform CFD
simulations with the LJTS model fluid.  Hence, we restrict ourselves in the following to the simpler PeTS EOS. A Fortran implementation of the PeTS EOS is being published in conjunction with this
paper, cf. \cite{petsdata}.

\subsection{Two-Phase Riemann Problem}

In the sharp interface method, we establish the coupling between the bulk phases by solving the Riemann problem. In general, the Riemann problem is an initial value problem for a one-dimensional
system of evolution equations with piecewise constant  initial data
\begin{equation}
\U \l(x,t=0\r) = 
\begin{cases}
\U_{\liq} \quad & \text{for} \quad x < x_0 \\
\U_{\vap} \quad & \text{for} \quad x > x_0
.
\end{cases}
\label{eq:multiphase_riemannproblem}
\end{equation}
For the Euler equations, the solution of the Riemann problem usually consists of four constant states, which are separated by simple waves. The outer waves are either a rarefaction wave or a shock
wave, while the intermediate wave is a contact discontinuity. For the ideal gas as well as for homogeneous real gases, the solution of the Riemann problem is well-known
\cite{godlewski_numerical_2014,toro_riemann_2009,colella_efficient_1985}.
 
In this work, we consider the Riemann problem, for which the initial states are in the vapor and in the liquid region, respectively. If phase transition occurs, a fourth wave appears, which resembles
the interface. This can only happen if the classical theory of the Riemann problem solution breaks down. The Euler equations are no longer purely hyperbolic because  non-convex EOS do not guarantee
real eigenvalues. In the unstable region between the spinodals, the eigenvalues are imaginary. 
\begin{figure}[!ht]
   \centering
   \includegraphics[width=0.8\linewidth]{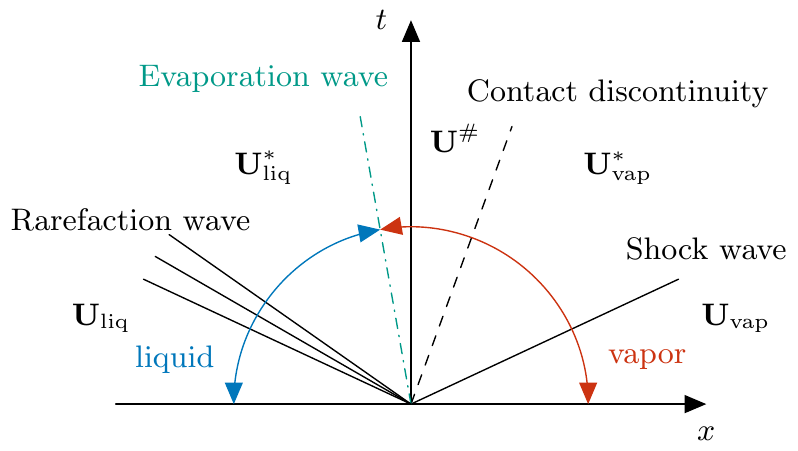}
   \caption{Solution structure for the two-phase Riemann problem of the two-phase shock tube scenario.}
  \label{fig:multiphase_sodproblem}
\end{figure}

In \cref{fig:multiphase_sodproblem}, the solution structure of a two-phase shock tube problem with evaporation is shown. The case with condensation was not considered in this study.  Neglecting heat
conduction at this point, the solution consists of five constant states, separated by four waves.  The outer waves are classical shock or rarefaction waves.  In between are a contact discontinuity and
the interface, where the latter is introduced as an undercompressive shock wave.  In the exemplary two-phase shock tube problem in \cref{fig:multiphase_sodproblem}, the liquid phase $\U_{\liq}$
expands rapidly due to a rarefaction wave until a metastable state $\U_{\liq}^*$ is reached.  This state undergoes phase transition across an undercompressive shock and a state $\U^\#$ appears,
constituted of freshly evaporated vapor.  A shock wave then propagates through the vapor phase.  The contact discontinuity is the material boundary between the freshly evaporated vapor $\U^\#$ and the
compressed vapor $\U^*_{\vap}$. Shock, rarefaction and contact waves propagate through homogeneous phases only. Hence, the underlying EOS remains convex and the waves are classical waves as in a
single phase Riemann problem.

\subsubsection{Evaporation Wave}
\label{sec:evaporationwave}

Special care has to be taken to model the evaporation wave as an undercompressive shock.  Across this non-classical wave, the jump conditions \cref{eq:jump_1,eq:jump_2,eq:jump_3} apply.  Moreover, to
obtain a unique solution, the entropy condition (\ref{eq:entropycondition_jump}) has to be fulfilled.  In contrast to a classical shock wave, the Lax or Liu entropy criteria are not sufficient.  We
therefore follow Refs. \cite{rohde_riemann_2018,fechter_sharp_2017} and rely on the concept of the kinetic relation.  It can be understood as a jump condition that provides the mass flux across an
undercompressive shock and accounts explicitly for the entropy production due to phase change. The difficulty is that the entropy production needs information from the microscale at the interface.  In
this work, we extended the framework of Rohde and Zeiler \cite{rohde_relaxation_2015,rohde_riemann_2018} from the isothermal Euler equations to the Euler-Fourier equations, following the theory of
classical non-equilibrium thermodynamics \cite{de_groot_non-equilibrium_1984}. A more detailed description is given in Ref. \cite{hitz2020}.   

We write the kinetic relation in the form 
\begin{equation}
\label{eq:kineticrelation}
\mathcal{K}:= f_m J_m+f_e J_e - \eta_{\Gamma} ( \dot{m}, J_{e} ) = 0,
\end{equation}
where $J_m=\dot{m}$ is the mass flux, $J_e$ the energy flux and $f_m$, $f_e$ the corresponding driving forces. The energy jump condition (\ref{eq:jump_3}) can be combined with the entropy jump condition
\eqref{eq:entropycondition_jump} to obtain the entropy condition in the form   
\begin{equation}
\label{eq:entropy_condition_fluxforceformulation}
- \dot{m} \llbracket \frac{g}{T} + \half \frac{\l( \vninterface - S \r)^2}{T} \rrbracket
+ J_{e} \llbracket \oneover{T} \rrbracket 
= \eta_{\Gamma}
,
\end{equation}
where enthalpy $h=\epsilon+p/\rho$ and Gibbs energy $g=h-Ts$ are introduced and the total energy flux across the interface is defined by
\begin{equation}
\label{eq:energyflux}
J_{e} = \qninterface + \dot{m} \l( h + \half \l( \vninterface - S \r)^2 \r)
.
\end{equation}
Macroscopic entropy production can therefore be expressed as a sum of flux/driving force pairs 
\begin{equation}
\label{eq:macroscopic_entropy_production}
f_m J_m+f_e J_e = - \dot{m} \llbracket \frac{g}{T} + \half \frac{\l( \vninterface - S \r)^2}{T} \rrbracket
+ J_{e} \llbracket \oneover{T} \rrbracket 
.
\end{equation}
Using these flux/driving force pairs, constitutive laws can be derived as a model for the entropy production on the microscale.  The classical approach \cite{de_groot_non-equilibrium_1984} assumes a
linear relation between fluxes and their driving forces
\begin{align}
\label{eq:entropycondition_linearrelation_mass}
- \llbracket \frac{g}{T} + \half \frac{\l( \vninterface - S \r)^2}{T} \rrbracket = \Rnn \dot{m} \quad & \text{with} \quad \Rnn(\Tinterface) \geq 0, \\
\label{eq:entropycondition_linearrelation_energy}
\llbracket \oneover{T} \rrbracket  = \Rqq J_{e} \quad & \text{with} \quad \Rqq(\Tinterface) \geq 0
.
\end{align}
The material parameters $\Rnn,\Rqq$ are called transport resistivities against mass and energy flux, respectively.  They are linear and do not depend on their respective driving force, but they do
depend on local state variables such as the interface temperature $\Tinterface$ \cite{kjelstrup_non-equilibrium_2008, heinen_communication:_2016}. It is bounded by the temperatures on both sides of
the interface, i.e. $\min\l(T_{\liq},T_{\vap}\r) \leq \Tinterface < \max\l(T_{\liq},T_{\vap}\r)$, and is unique, i.e. $\llbracket \Tinterface \rrbracket \equiv 0$.  The resistivities are well-known
from Onsager theory \cite{kjelstrup_non-equilibrium_2008} and describe entropy producing processes on the microscale.  They can be obtained e.g. with density functional theory (DFT) methods
\cite{johannessen_nonequilibrium_2008,klink_analysis_2015} or molecular dynamics methods \cite{Simon2006}.

With the constitutive expressions for the flux/driving force pairs, entropy production on the microscale can be expressed as
\begin{equation}
\label{eq:entropy_production_squaresum}
\eta_{\Gamma}( \dot{m}, J_{e}) = \Rnn \dot{m}^2 + \Rqq J_{e}^2,
\end{equation}
neglecting cross correlations between the flux/driving force pairs.  The full kinetic relation reads
\begin{equation}
\label{eq:kineticrelation_specific}
\mathcal{K}_{\dot{m},e}:  
- \dot{m} \llbracket \frac{g}{T} + \half \frac{\l( v - S \r)^2}{T} \rrbracket
+ J_{e} \llbracket \oneover{T} \rrbracket 
- \Rnn \dot{m}^2 - \Rqq J_{e}^2
= 0
.
\end{equation}
Unfortunately, this relation is highly non-linear since the energy flux is not only determined by the heat transfer across the phase interface, but also depends on the mass flux, cf.
\cref{eq:energyflux}.  This makes the analysis of the kinetic relation complicated and the solution procedure for the two-phase Riemann problem difficult. Hence, we simplified the relation in the present
simulations by neglecting the entropy production due to energy transport.  We also assumed that temperature jumps may appear in the solution of the macroscopic Riemann problem, but the entropy
production is determined at a unique interface temperature $\Tinterface$.  The kinetic relation employed in these simulations was therefore simplified to 
\begin{equation}
   \label{eq:kineticrelation_isothermal_finished}
\mathcal{K}:= - \dot{m} \llbracket g + \half \l( v - S \r)^2 \rrbracket  - \Tinterface \Rnn \dot{m}^2 = 0
 .
\end{equation}
It should be noted that the heat flux is required to construct the entropy condition across the interface.  While the isothermal Euler equations implicitly account for heat transfer, a fully adiabatic
system produces a contradiction between the energy jump condition and the material law.  \citet{hantke_impossibility_2019} considered vapor-liquid interfaces at subcritical temperatures for the
Euler equations, i.e. $\qninterface\equiv 0$.  In this case, ${\rho_{\vap}^{-1}}>{\rho_{\liq}^{-1}}$ holds and for a non-zero mass flux, the energy jump condition \eqref{eq:jump_3} becomes
\begin{equation}
\llbracket h \rrbracket = - \frac{\dot{m}^2}{2} \llbracket \rho^{-2} \rrbracket < 0
.
\end{equation}
However, due to the enthalpy of evaporation, the enthalpy of the vapor is larger than the enthalpy of the liquid so that the material law produces the contradiction 
\begin{equation}
\llbracket h \rrbracket >0 
.
\end{equation}
Hence, without consideration of a heat flux, it is impossible to model phase transition with the adiabatic Euler equations.

\subsubsection{Modelling Heat Transfer Across the Interface}
\label{sec:heat}

To model heat transfer across the interface, a microscale model was constructed on the basis of Fourier's law.  The approach rests on the ideas of Gassner \etal\cite{gassner_contribution_2007} and
L\"orcher \etal\cite{lorcher_explicit_2008} in their construction of a numerical diffusion flux. They considered the Riemann problem for the linearized heat equation with a jump of the thermal
diffusivity $k$ at an interface with the position $x=0$.  The solutions are obtained separately for each region with constant $k$ and are defined by $T^-$ and $T^+$, respectively. The different
constant thermal diffusivities are denoted by $k^+$ and $k^-$.  The heat equation with the discontinuous coefficient reads
\begin{equation}
\label{eq:heat_leftright}
   T^{\pm}_t - k^{\pm} T^{\pm}_{xx} = 0
,
\end{equation}
where
\begin{equation}
k^{\pm} = \frac{\lambda^{\pm}}{\rho^{\pm} c_v^{\pm}}
.
\end{equation}
At $x=0$, compatibility conditions  
\begin{align}
\label{eq:heat_coupling_1}
T^+(0,t) & = T^-(0,t), \\
\label{eq:heat_coupling_2}
k^+ T^+_x(0,t) & = k^- T^-_x(0,t),
\end{align}
are imposed such that temperature and flux of internal energy are continuous.

Following Refs. \cite{gassner_contribution_2007,lorcher_explicit_2008}, we consider the diffusive generalized Riemann problem (dGRP) for \cref{eq:heat_leftright} with the initial data 
\begin{equation}
\label{eq:dgrp_initial}
T (x,t=0) = 
\begin{cases}
T_{L} \quad & \text{if} \quad x < 0, \\
T_{R} \quad & \text{if} \quad x > 0. \\
\end{cases}
\end{equation}
The subgrid model computes the heat fluxes on both sides of the interface based on a time averaged solution of the dGRP with initial data corresponding to the
temperatures of the inner states $\U_{\liq}^*$ and $\U^{\#}$ of the solution of the two-phase Riemann problem as shown in \cref{fig:multiphase_sodproblem}.

\begin{figure}[!ht]
   \centering
   \includegraphics[width=0.5\linewidth]{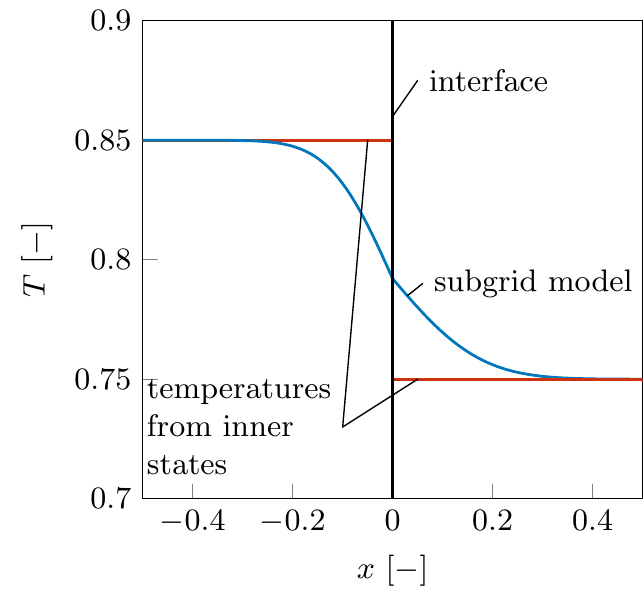}
   \caption{Modelling heat transfer across the phase interface using a subgrid model to link the temperature from the inner states of the solution of the two-phase Riemann problem.}
   \label{fig:heattransfer_subgridmodel}
\end{figure}

An exact solution of Eqs. (\ref{eq:heat_leftright}) and (\ref{eq:dgrp_initial}) with the conditions (\ref{eq:heat_coupling_1}) and (\ref{eq:heat_coupling_2}) can be determined by Laplace
transformation \cite{gassner_contribution_2007,lorcher_explicit_2008}. This approach provides the opportunity to define the interface temperature by averaging over the time interval $[0,\Delta
t]$
\begin{equation}
\label{eq:heat_solution_interface_average}
\Tinterface = \oneover{\Delta t} \int_0^{\Delta t} T(0,t) \d t = T_{L} + 
\frac{\l( T_{R}-T_{L} \r) \sqrt{k^+}}{ \l( \sqrt{k^+} + \sqrt{k^-} \r)}
.
\end{equation}
The thermal flux is singular for $t=0$, but the integration over time can be expressed as an improper integral by  time  averaging over the interval $[0,\Delta t]$
\begin{equation}
\label{eq:heat_flux_interface_average}
g_a =  
\frac{ 2 \l( T_{R}-T_{L} \r) \sqrt{k^+k^-}}{ \sqrt{\pi \Delta t} \l( \sqrt{k^+} + \sqrt{k^-} \r)} 
.
\end{equation}
The values of the heat flux on both sides of the interface can then be evaluated as 
\begin{align}
\label{eq:heat_heatflux_left}
q^- & = \rho^- c_{v}^- g_a, \\
\label{eq:heat_heatflux_right}
q^+ & = \rho^+ c_{v}^+ g_a
.
\end{align}
A solution of the microscale model for piecewise constant initial data is shown in \cref{fig:heattransfer_subgridmodel} at the time instance $t=1.0$.

\subsection{Homogeneous Equilibrium Model}
\label{sec:hem}
In addition to this non-equilibrium model, we also consider the HEM.  Its use in numerical simulations was introduced by \citet{stewart_two-phase_1984}. In this context, discussions about the solution
of the Riemann Problem can be found in Ref. \cite{menikoff_riemann_1989}. For a pure fluid, HEM is based on the assumption of thermodynamic equilibrium in the two-phase region 
\begin{equation}
\label{eq:equilibriumcondition}
T_{\vap} = T_{\liq}, \quad
p_{\vap} = p_{\liq}, \quad
g_{\vap} = g_{\liq}
.
\end{equation}
The assumption of equilibrium conditions implies that two-phase states only exist as a mixture of saturated liquid and saturated vapor, i.e. as a hypothetical \textit{homogeneous} wet vapor. This
avoids the consideration of metastable or unstable states.  Dynamic phase change processes are assumed to occur so fast that temperature and pressure in the control volume are immediately uniform due
to rapid acoustic pressure waves and heat conduction. In the two-phase region, the wet vapor can be described by the vapor quality
\begin{equation}
\label{eq:hem_vapour_quality}
q_{\vap} = \frac{m_{\vap}^{\sat}}{m} ,
\end{equation}
with the mass of saturated vapor $m_{\vap}^{\sat}$ and the overall mass $m$.

\begin{figure}[!ht]
   \centering
   \includegraphics[width=0.8\linewidth]{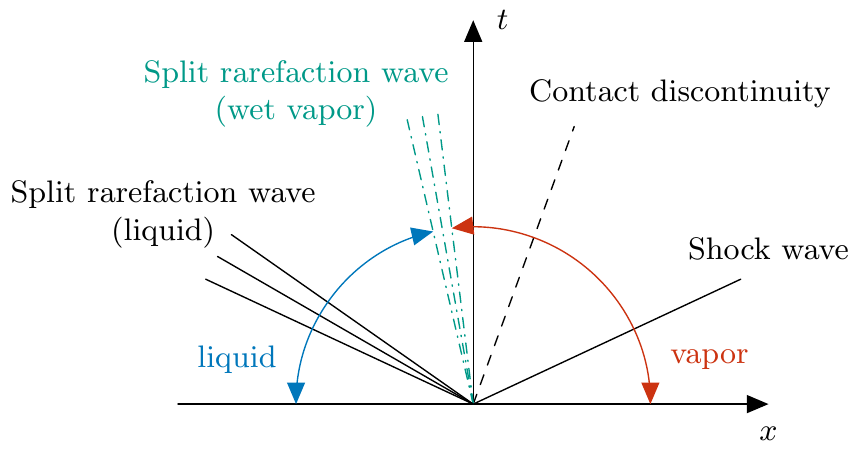}
   \caption{Solution structure of the two-phase Riemann problem for a two-phase shock tube scenario with the HEM approach.}
  \label{fig:multiphase_hem}
\end{figure}
The Maxwell line denotes the path of thermodynamic equilibrium in the two-phase region, cf. Fig. \ref{fig:ljts_isothermal}. Along this path, phase transition is considered isothermal and isobaric, as
well as maintaining a constant Gibbs energy.  If the fluid is in the two-phase region, the thermal EOS is replaced by the equilibrium conditions  (\ref{eq:equilibriumcondition}) and the caloric EOS is
replaced by a convex combination of the internal energies of the saturated phases.  Due to this crossover at the binodal, pressure and entropy exhibit kinks along the isotherm at saturation conditions
and a direct differentiation to calculate the speed of sound fails.  A common approach to model the speed of sound of the wet vapor was introduced by Wood \cite{wood_textbook_1930}. Due to the
different speeds of sound in the homogeneous phases and the wet vapor,  non-classical wave behavior occurs. The classical rarefaction wave is replaced by so-called split waves as depicted in
\cref{fig:multiphase_hem}. A more detailed description of the HEM approach can be found in Ref. \cite{foll_use_2019}. 

\section{Numerical Methods}
\label{sec:numericalmethods}

\subsection{Fluid-Solver FLEXI}

The conservation equations were solved with the DGSEM,  which is implemented in the open source code \flexi\footnote{https://www.flexi-project.org/}. The numerical method was described in detail by
Hindenlang et al. \cite{hindenlang_explicit_2012}. Hence, we only give a short overview of the basic building blocks to convey an impression of this numerical method. As usual in the discontinuous
Galerkin approach,  the solution and the fluxes are approximated by polynomials in each grid element allowing for discontinuities between the elements. In DGSEM, the polynomial basis is a nodal one
with Lagrangian polynomials defined by Gauss points.  The physical grid cell is projected to a reference element and a weak formulation of the conservation equations is derived in this reference
element.  The resulting volume and surface integrals are approximated by Gauss quadratures.  As for FV schemes, numerical flux functions are needed to couple the neighboring elements.  If
discontinuities, such as shock waves or material boundaries, appeared in the solution, a shock capturing method \cite{sonntag_shock_2014, sonntag_efficient_2017} was activated, in which the elements
with oscillations are refined by sub-cells, and the numerical method switched locally to a total variation diminishing FV method on the sub-cells. The number of sub-cells coincides with the number of
degrees of freedom. The indicator of \citet{persson_sub-cell_2006} was used to detect oscillations. For time integration, an explicit high order Runge-Kutta scheme \cite{kennedy_low-storage_2000} was
used.

\subsection{Solution of the Two-Phase Riemann Problem}

The solution strategy proposed here is an extension of the method of Fechter \etal\cite{fechter_sharp_2017}.  In the present work, heat transfer was taken into account with Fourier's law. In this
case, the self-similarity of the Riemann solution breaks down. It is no longer a sole function of $x/t$, but depends explicitly on time and space. To circumvent this problem, we solve the heat
conduction only at the interface as discussed in Section \ref{sec:heat}. From the subgrid model, we  obtain the interface  temperature and the time averaged heat flux for the interval $[0,\Delta t]$
on both sides and use them in the jump conditions across the interface.  Hence, this "microscale" information affects only the jump conditions across the interface and the classical hyperbolic
solution procedure can be used for the Riemann solver. 

The solution consists of two outer elementary waves which are either classical shock or rarefaction waves.  An undercompressive shock wave represents the interface and a contact discontinuity
provides the material boundary between the compressed vapor and the freshly evaporated vapor.  Classical shock waves satisfy the Rankine-Hugoniot conditions, while the Riemann invariants
\cite{godlewski_numerical_2014} are constant across the rarefaction waves. The contact discontinuity is in mechanical equilibrium, and across the interface the jump conditions
\eqref{eq:jump_1} to \eqref{eq:jump_3} hold with the additional constraint of the kinetic relation \eqref{eq:kineticrelation_isothermal_finished}. The resulting system of equations is highly
non-linear and an iterative scheme was used to solve it.

\begin{figure}[!ht]
   \centering
   \includegraphics[width=0.8\linewidth]{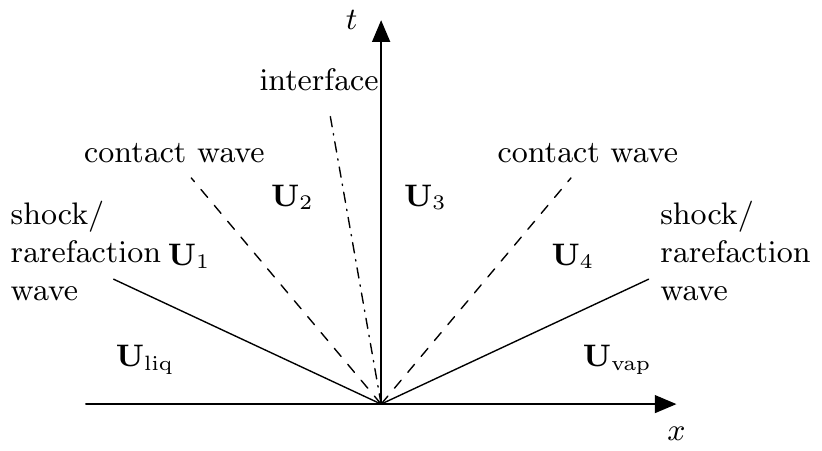}
   \caption{Modified solution structure of the two-phase Riemann problem with a liquid state  and a vapor state.}
  \label{fig:multiphase_solutionmethod}
\end{figure}

When the state of aggregation of $\U^\#$ changes during iterations, the scheme may fail because the densities and entropies of the liquid and vapor differ largely from each other.  To alleviate this
numerical problem, similar to Ref. \cite{fechter_sharp_2017}, a second contact wave was introduced as shown in \cref{fig:multiphase_solutionmethod}.
The solution now contains four intermediate states where $\U_1$ and $\U_2$ are in the liquid phase and $\U_3$ and $\U_4$ in the vapor phase.  Once the solution structure was known, the second contact
wave was removed. 

The target function of the iteration scheme was defined as
\begin{equation}
\label{eq:multiphase_targetfunction}
G_{\mathrm{MRP}} \l( \tau_1,T_1,\tau_2,T_2,\tau_3,T_3,\tau_4,T_4 \r) = \l(r_1,r_2,r_3,r_4,r_5,r_6,r_7,r_8 \r)\transpose = 0
.
\end{equation}
It minimizes the residuals $r_{i}$ for the specific volumes and temperatures of the states $\U_1$, $\U_2$, $\U_3$ and $\U_4$ as well as the initial states $\U_{\liq}$ and $\U_{\vap}$. The solution
procedure was as follows.
\paragraph{Step 1: Initial guess} 
As an initial guess for the iteration procedure, the states on both sides of the interface are assigned by the corresponding saturation conditions at the given temperature
\begin{equation}
\label{eq:multiphase_initialguess_sat}
\U_1 = \U_2 = \U_{\liq}^{\sat}(T=T_{\liq}), \quad \text{and} \quad \U_3 = \U_4 = \U_{\vap}^{\sat}(T=T_{\vap})
.
\end{equation}
For each state, the following thermodynamic properties are calculated with the PeTS EOS
\begin{align}
\label{eq:multiphase_initialguess_calcvars}
& p_i = p(\tau_i,T_i), \quad
\epsilon_i = \epsilon(\tau_i,T_i), \quad
h_i = h(\tau_i,T_i), \\
& s_i = s(\tau_i,T_i), \quad
g_i = g(\tau_i,T_i).
\end{align}
\paragraph{Step 2: Residuals for the outer waves}
The outer waves are classical shock or rarefaction waves.  For $p_1<p_{\liq}$, the wave in the liquid is a rarefaction wave, where the Riemann invariants are satisfied, such that an equation for the
velocity is given with the specific volume $\tau$ as the working variable.  The corresponding residual is the constant entropy
\begin{align}
\label{eq:multiphase_solution_rare_left_v}
      v_1 & = v_{\liq} + \int_{\tau_{\liq}}^{\tau_1}\frac{c}{\tau} \d \tau
      \quad \text{with} \quad
      c=c\l(\tau,s_{\liq}\r) , \\
\label{eq:multiphase_solution_rare_left_residual}
      r_1 & = s_1 - s_{\liq}
      ,
\end{align}
with $c$ being the speed of sound. For $p_1\geq p_{\liq}$, the wave in the liquid is a shock wave.  The post shock velocity is calculated from the Rankine-Hugoniot conditions and the residual is the
Hugoniot equation
\begin{align}
\label{eq:multiphase_solution_shock_left_v}
v_1 & = v_{\liq} - \sqrt{\l|\l(p_{\liq}-p_1\r)\l(\tau_1-\tau_{\liq}\r)\r|} , \\
\label{eq:multiphase_solution_shock_left_residual}
r_1 & = \epsilon_1 - \epsilon_{\liq} + \half \l(p_{\liq}+p_1\r) \l(\tau_1-\tau_{\liq}\r)
.
\end{align}
Analogously, for  $p_4<p_{\vap}$, the outer wave in the vapor is a rarefaction wave
\begin{align}
\label{eq:multiphase_solution_rare_right_v}
      v_4 & = v_{\vap} - \int_{\tau_{\vap}}^{\tau_4}\frac{c}{\tau} \d \tau
      \quad \text{with} \quad
      c=c\l(\tau,s_{\vap}\r) , \\
\label{eq:multiphase_solution_rare_right_residual}
      r_2 & = s_4 - s_{\vap}
      ,
\end{align}
and for $p_4\geq p_{\vap}$, the outer wave in the vapor is a shock wave
\begin{align}
\label{eq:multiphase_solution_shock_right_v}
v_4 & = v_{\vap} + \sqrt{\l|\l(p_{4}-p_{\vap}\r)\l(\tau_{\vap}-\tau_{4}\r)\r|} , \\
\label{eq:multiphase_solution_shock_right_residual}
r_2 & = \epsilon_{\vap} - \epsilon_{4} + \half \l(p_{4}+p_{\vap}\r) \l(\tau_{\vap}-\tau_{4}\r)
.
\end{align}
\paragraph{Step 3: Residuals for contact waves}
Across both contact waves, velocities and pressures are uniform
\begin{align}
\label{eq:multiphase_solution_contact_v}
v_2 = v_1; \quad & v_3=v_4 , \\
\label{eq:multiphase_solution_contact_residuals}
r_3 = p_2-p_1; \quad & r_4 = p_4-p_3
.
\end{align}
\paragraph{Step 4: Phase transition mass flux}
The mass flux across the interface and the corresponding wave speed are
\begin{equation}
\label{eq:multiphase_solution_massflux_speed}
\dot{m} = \frac{v_3-v_2}{\tau_3-\tau_2}; \quad S^\# = v_2 - \dot{m} \tau_2 = v_3 - \dot{m} \tau_3
.
\end{equation}
\paragraph{Step 5: Remove one of the two contact waves}
Once the phase transition speed $S^\#$ and the propagation speeds $v_{3,4}$ of the contact waves are determined, the wave pattern is known and the redundant contact wave can be removed
\begin{equation}
\label{eq:multiphase_solution_erase_contact}
r_5 = \begin{cases}
T_4 - T_3 \quad & \mathrm{if} \quad \dot{m}<0, \\
\l(T_2-T_1\r)\l(T_4-T_3\r) \quad & \mathrm{if} \quad \dot{m}=0, \\
T_2 - T_1 \quad & \mathrm{if} \quad \dot{m}>0
.
\end{cases}
\end{equation}
\paragraph{Step 6: Heat flux}
The heat fluxes $q^+$ and $q^-$ are calculated by the microscale model
\begin{align}
\label{eq:multiphase_solution_heatflux_left}
q^{-} & = \rho_2 c_{v,2} g_a, \\
\label{eq:multiphase_solution_heatflux_right}
q^{+} & = \rho_3 c_{v,3} g_a
,
\end{align}
with the initial data $T^-(t=0)=T_2$ and $T^+(t=0)=T_3$.
\paragraph{Step 7: Jump conditions across the interface}
Finally, the jump conditions \eqref{eq:jump_1} to \eqref{eq:jump_3} and the kinetic relation \eqref{eq:kineticrelation_isothermal_finished} need to be fulfilled across the interface
\begin{align}
\label{eq:multiphase_solution_phase_residual_momentum}
r_6 & = \dot{m} \l( v_3-v_2 \r) + p_3-p_2 - \surfacetension , \\
\label{eq:multiphase_solution_phase_residual_energy}
r_7 & = \dot{m} \l( h_3 - h_2 + \half \dot{m}^2 \l(\tau_3-\tau_2\r)\r)  + q_+ - q_- , \\
\label{eq:multiphase_solution_phase_residual_kineticrelation}
r_8 & = \mathcal{K}_{\dot{m}}
.
\end{align}

The two-phase Riemann problem is then solved for the initial data $\U_{\liq}$, $\U_{\vap}$ and constant resistivity $\Rnn$.  First, the temperatures $T_{\liq}$ and $T_{\vap}$ are calculated and the
initial guess is assigned.  Then the residuals $r_1,\ldots,r_8$ are computed, alongside velocities $v_1,\ldots,v_4$ and mass flux $\dot{m}$, following the steps discussed above.  The equation
$G_{\mathrm{MRP}}=0$ is then solved numerically by an eight-dimensional root finding algorithm provided by the open source libraries GSL (V2.1) and FGSL (V1.2.0).

\subsection{Sharp Interface Ghost Fluid Method}
Numerical methods for sharp interface models require additional strategies to track the interface and to impose the two-phase jump conditions.  For this study, the method of Fechter
\etal\cite{fechter_compressible_2015,fechter_sharp_2017,fechter_approximate_2018} was simplified to a one-dimensional front-tracking scheme.  In the bulk phases, the solution was obtained by the DGSEM
solver \flexi.

The interface is marked by the zero of a level-set function $\Phi$. It is advected with a color function \cite{sussman_level_1994}
\begin{equation}
\label{eq:numerics_ft_levelset}
(\Phi)_t + S \cdot \gradient \Phi = 0
,
\end{equation}
where $S=S(t) \in \R$ is the propagation speed of the interface. Because $S_x \equiv 0$  in one spatial dimension, \cref{eq:numerics_ft_levelset} can be reformulated as a conservation
equation
\begin{equation}
\label{eq:numerics_ft_levelset_conservativeform}
(\Phi)_t + \div \l( S \Phi \r) = 0
,
\end{equation}
which can be solved straightforwardly by DGSEM.

Initially, the interface location is positioned on an element face.  To keep the interface sharp, the mesh is advected by the interface speed using the ALE method \cite{etienne_geometric_2008}. The
interface speed is given by the solution of the two-phase Riemann problem.  The mesh velocity then only depends on time and is constant in space. Thus, the geometric conservation law is always
fulfilled discretely.  A description of the ALE method in the context of DGSEM was given by Minoli and Kopriva \cite{minoli_discontinuous_2011}.  The ALE contribution to the fluxes in the volume
integral was taken into account and the numerical flux function was modified similar to Ref. \cite{shen_robust_2014}.

\begin{figure}[htb] 
   \centering
   \includegraphics[width=0.49\linewidth]{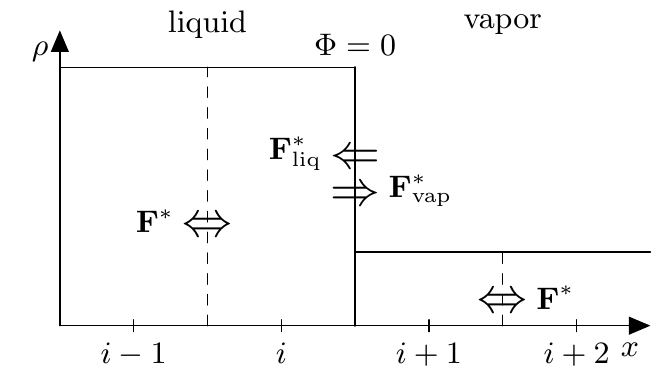} 
   \caption{ Calculation of discontinuous fluxes at the interface.  Both fluxes are obtained from the
solution of the two-phase Riemann problem and account for the jump conditions.  } 
   \label{fig:numerics_gfm} 
\end{figure}

Across the interface, the two-phase jump conditions are fulfilled by the solution of the two-phase Riemann problem. We follow the ideas of a ghost fluid approach as proposed in Refs.
\cite{fechter_compressible_2015,fechter_sharp_2017,fechter_approximate_2018,merkle_sharp-interface_2007} and use this solution to define the ghost states. The inner states at the interface
$\U^*_{\liq}$ and $\U^\#$ are taken as ghost states for the respective bulk phase. This guarantees that proper waves in the liquid and vapor are generated.  The physical fluxes for the liquid and
vapor are then calculated from  $\Fb_{\liq}^*=\Fb(\U^*_{\liq})-\U^*_{\liq} S$ and $\Fb_{\vap}^*=\Fb(\U^\#)-\U^\# S$, where $\U S$ denotes the ALE contribution.  A schematic of this situation is shown in
\cref{fig:numerics_gfm}. At single phase element faces in the bulk, a unique flux is calculated by a numerical flux function.

\subsection{Molecular Dynamics Simulation}
\label{sec:md}

Molecular dynamics simulation provides a physically sound approach to investigate phase change processes because it rests on statistical mechanics. Except for the force field describing the intermolecular interactions as well as initial and boundary conditions, practically no further assumptions have to be made.

As we have done in the first part of this paper series \cite{hitz_comparison_2020}, molecular dynamics was employed to generate data sets serving as a benchmark for the numerical results of the
present macroscopic solution. The studied transient process is accompanied by rapid changes of local quantities. To nonetheless obtain simulation data with a good statistical quality, comparatively
large systems were sampled, containing up to $3\cdot 10^8$ particles. To cope with such a large number of particles, all simulations were carried out with the massively-parallel code \textit{ls1
mardyn} \cite{niethammer_ls1_2014}. This continuously developed code was recently further improved with respect to its node-level performance and parallel efficiency when executed on multiple compute
nodes \cite{twetris_2019}.

\begin{figure}[htb]
   \centering
   \includegraphics[width=0.69\linewidth]{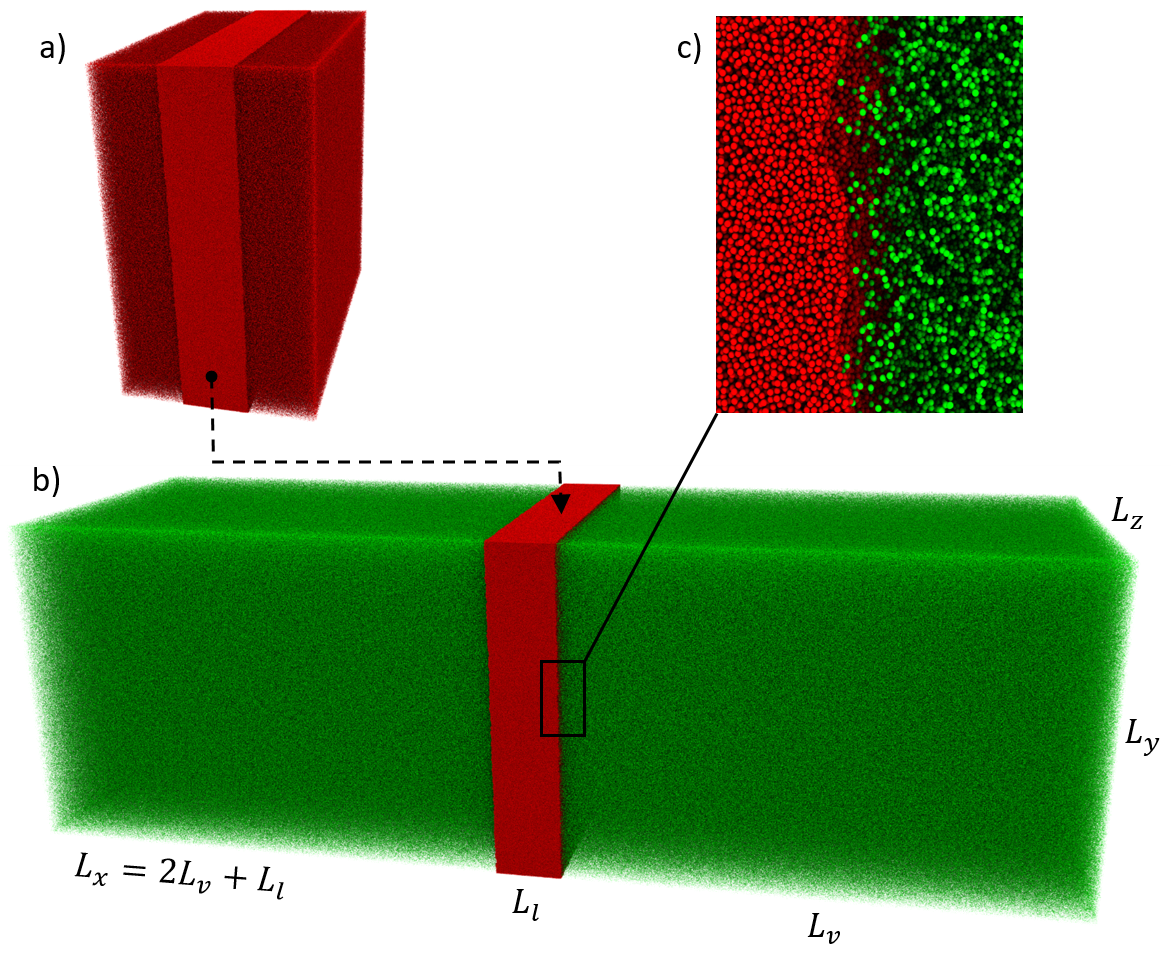}
   \caption{
Snapshots of the prepared molecular systems, rendered with the cross-platform visualization prototyping framework MegaMol \cite{megamol_2015} and the \textit{Intel OSPRAY}
plugin \cite{Gralka2019Megamol,rau17ospray}. a) Final configuration of the vapor-liquid equilibrium simulation from which the liquid phase was extracted to build the test case scenarios. b) One of the
test case scenarios with a vapor phase (green) diluted to 70\% of the saturated vapor density at the temperature $T=0.8$. c) Close-up look at the interface.}
   \label{fig:numerics_md}
\end{figure}

The particle ensemble's initial configuration for all considered test cases was prepared in two steps. First, a vapor-liquid equilibrium was simulated in the direct classical way: a planar film
containing a saturated liquid phase was brought into contact with a saturated vapor phase, yielding a symmetrical cuboid system as shown in \cref{fig:numerics_md}.
Initial values for the saturated vapor and liquid densities were taken from Ref. \cite{thol_equation_2015}. Both the liquid and vapor phases had an extent of $200$ particle diameters $\sigma$ and the
system's cross-sectional area was $10^6 \sigma^2$. To mimic an infinitely extended liquid film, periodic boundary conditions were imposed in all spatial directions. The liquid and vapor
phases were thermostated by dividing the system into bins along the $x$ direction, each covering a range of $10\sigma$, and a constant temperature was maintained independently inside each of
them during equilibration by velocity scaling.

To save computational resources, the vapor-liquid equilibrium simulations were in fact conducted initially with a substantially smaller cross-sectional area of $4\cdot 10^4 \sigma ^2$ for a sufficiently
long period of $5\times 10^5$ time steps. Subsequently, the system was scaled up by replicating it $5 \times 5$ times in the $yz$ plane and then equilibrated for another $5\times 10^4$ time steps
until the replication pattern had vanished.

The second step was to replace the saturated vapor phase with a gas phase under thermodynamic conditions that were intended for the test cases. Therefore, the liquid phase particles were extracted by
evaluating the local density that they experience within a sphere with a radius of $2.5 \sigma$ around every single particle. Particles experiencing a local density of less than half of the saturated
liquid density were rejected. The extracted liquid phase was then brought into contact with an equilibrated vapor phase under thermodynamic conditions that were intended for the test cases,
i.e. having densities of $50\%$, $70\%$ and $90\%$ of the saturated density at a temperature of $T=0.8$, cf. \cref{fig:numerics_md} b). Technically, the vapor phase particles were treated as another
pseudo-species to gain access to additional information about the interactions between the phases by sampling partial quantities, e.g. partial density profiles.

\begin{figure}[h!t]
   \centering
   \includegraphics[width=\linewidth]{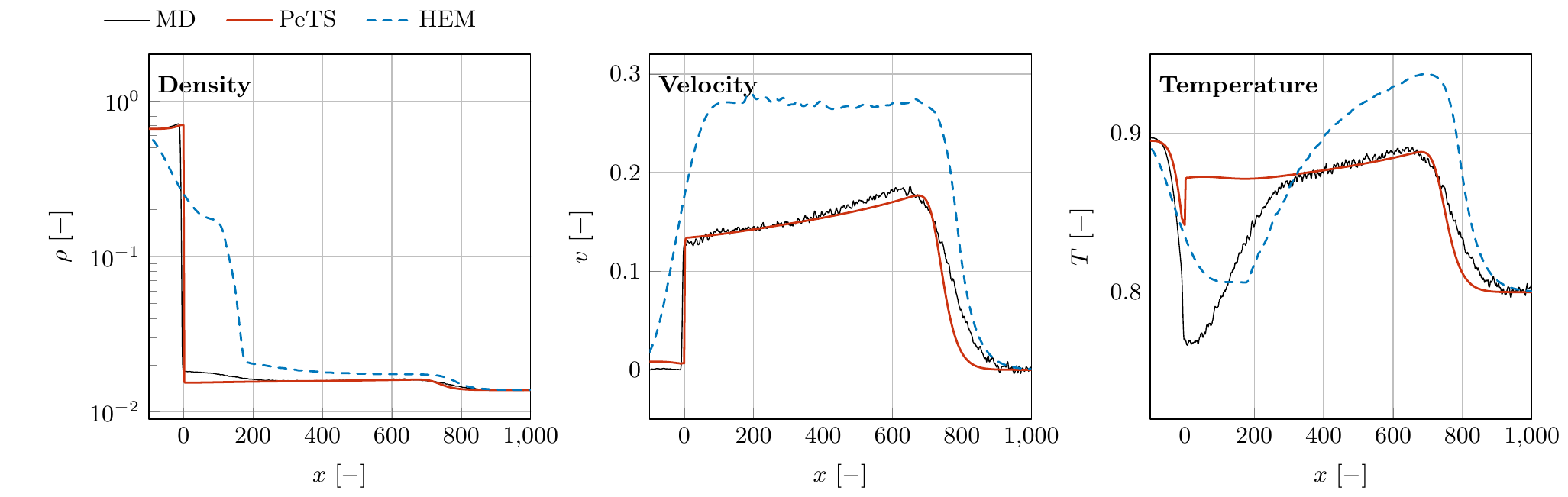}
   \caption{Molecular dynamics results for test cases 2 and 4 with different liquid film thickness. The plot shows the results for density, velocity and temperature at $t=600$.}
   \label{fig:results_liquidfilm}
\end{figure}

As revealed by the simulation results presented in Section 4, shock waves emerged from the two interfaces as a consequence of the global non-equilibri\-um, cf.  \cref{fig:results_liquidfilm}.
Because of the systems' symmetry and periodicity, the shock waves exerted from the two opposite interfaces meet each other when crossing the periodic boundary. At this time instance, sampling
results lose their value so that simulations were terminated. To nonetheless cover a sufficient amount of time, the vapor phases had an extent of $1500 \sigma$  in $x$ direction, such that
a period of at least $5\cdot 10^5$ time steps could be sampled. All system dimensions and initial liquid and vapor densities are summarized in \cref{tbl:md}.
\begin{table}[ht]
\caption{System dimensions and initial liquid and vapor densities of the test case scenarios.}
\centering
\begin{tabular}{l|c|c|c|c|c|c}
\toprule
 & $L_{\liq}$ & $L_{\vap}$ & $L_z$ & $\rho_{\liq}$ & $\rho_{\vap}$ & $N / 10^8$  \\
\midrule
Case 1 & 200 & 1500 & 3200 & 0.6635 & 0.017800 & 1.61   \\
Case 2 & 200 & 1500 & 3200 & 0.6635 & 0.013844 & 1.73   \\
Case 3 & 200 & 1500 & 3200 & 0.6635 & 0.009889 & 1.85   \\
Case 4 & 400 & 1500 & 3400 & 0.6635 & 0.013844 & 3.07   \\
\bottomrule
\end{tabular}
\label{tbl:md}
\end{table}

Another issue that demanded attention was the evaporative cooling effect. Under phase equilibrium conditions, the particle fluxes of evaporating and condensing particles at the interface
are balanced.  Under non-equilibrium conditions, imposed by the vapor phase in the present test cases, however, a net flux of evaporating particles arose. Since particles with a high kinetic energy
are preferentially able to overcome the attractive force of the interface to propagate into the vapor phase, the slower ones remained, which entailed that the liquid cooled down and that its density
increased. This phenomenon initiated a motion of the liquid surface towards the center of the system. Again, owed to the system's symmetry, this motion was initiated at both interfaces, but in
opposing direction and hence these waves met in the system's center. Of course, this was an unwanted issue. Since the propagation of that motion was too fast, this issue could not be resolved by a
sufficiently wide liquid film because this would require an infeasible particle number. Instead, test case 2 was repeated as test case 4 with a doubled width of the liquid film, to assess the
influence of the delayed collision of the waves on the process evolving in the vapor phase. The results of both simulations are compared in \cref{fig:results_liquidfilm}. The excellent agreement
of the profiles indicates a negligible influence of the liquid film thickness on the properties of the vapor phase.

To capture the rapid dynamics of the system, a classical binning scheme with a high spatial and temporal resolution was employed to record the profiles of interest, i.e. temperature, density and
hydrodynamic velocity. These profiles were sampled within bins of $0.5\sigma$ width and averaged over a period of $500$ time steps. The number of time steps, i.e. particle configurations taken into
account for the average, was a compromise between suppressing atomistic fluctuations, without blurring the rapidly changing profiles. The specified spatial resolution turned out to be too extreme so
that the profiles were smoothed spatially by post-processing, without affecting their characteristic shape. A sliding window over ten adjoining data points was employed to calculate a running average.

\begin{figure}[h!t]
   \centering
   \includegraphics[width=\linewidth]{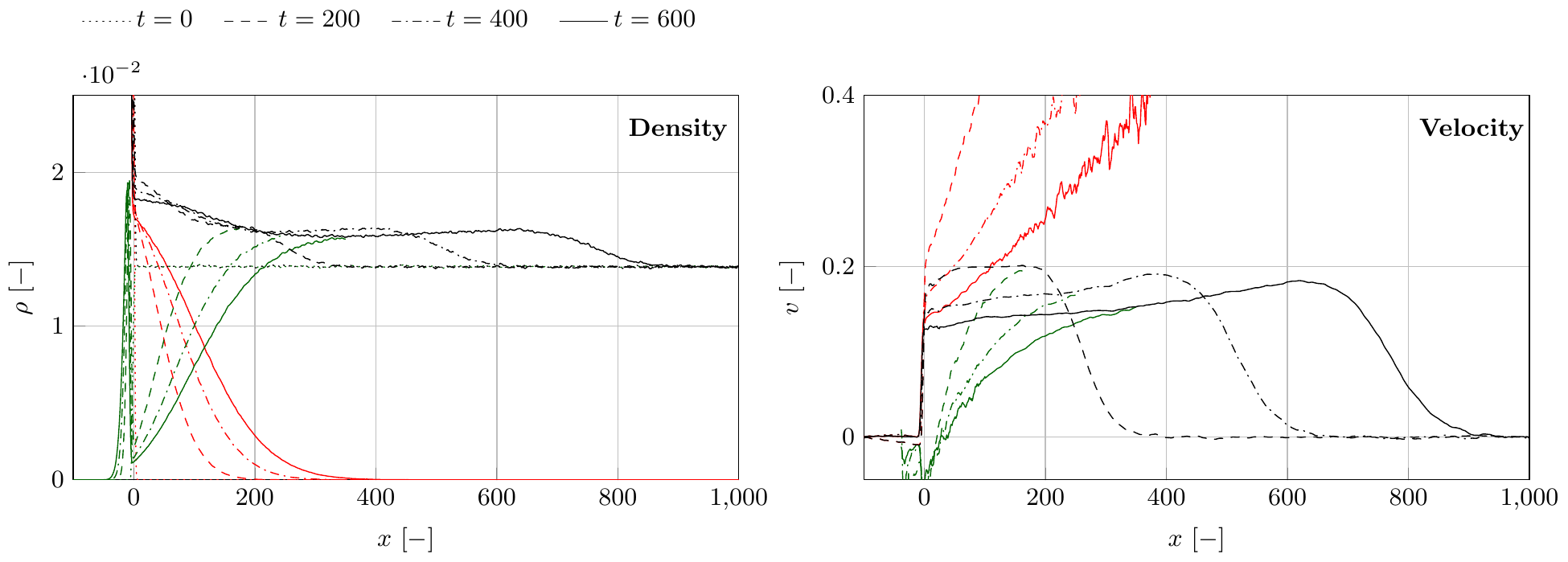}
   \caption{ Density (left) and velocity profiles (right) at four time instances $t = 0,\, 200,\, 400$ and $600$. Partial density and partial velocity profiles were sampled considering only particles that
   were initially constituting the liquid (red) or the vapor phase (green), respectively. The total density and velocity profiles (black) are the weighted sum of the partial ones.}
   \label{fig:particles}
\end{figure}

Treating the particles that constitute the liquid phase and the vapor phase in the initial configuration as distinguishable pseudo-species revealed additional insight into the process, cf.
\cref{fig:particles}.  For convenience, these pseudo-species are addressed in the following by “liquid particles” and “vapor particles”, respectively, although they may change their state of
aggregation over the course of time.

The partial density profiles show that when the shock had almost reached the system boundary $x=1000$ at the time instance $t=600$, the foremost liquid particles had propagated only half this
extent, although they obviously induced the shock wave to emerge. Approximately at half way between the interface and the propagation front of the liquid particles, an equimolar composition of the
pseudo-species, where partial density profiles cross, was observed. This point represents the material boundary, which smears out by diffusion and propagates away from the interface with the contact
wave speed, cf. \cref{fig:multiphase_sodproblem}. Directly at the interface, a jump in the partial density profile of the vapor particles was observed. It illustrates that vapor particles were
absorbed by the liquid, diffusing into it over time.

The partial velocity profiles show that the liquid particles on average move faster than the vapor particles, which is not surprising because of the higher initial temperature of the liquid phase.
However, the velocities assimilate over time, indicating momentum transfer from liquid to vapor particles. Approaching regions of low partial density, these profiles become
increasingly noisy due to statistical reasons. The total velocity of the vapor phase is the sum of the partial velocities, weighted by the local partial densities. Hence, the total velocity profile lies
in between the partial velocity profiles. Near the interface, it is dominated by the velocity of the liquid particles, while the velocity of the vapor particles dominates near the shock wave.  With
the progress of time, a decrease of the total velocity in the vicinity of the interface was observed. This can be explained by cooling due to evaporation as discussed above. 

\section{Results}
\label{sec:results}

Results are presented for a two-phase shock tube scenario with three different sets of initial conditions.  All physical properties were non-dimensional\-ized in terms of reference length
$\sigma_{\mathrm{ref}}=1\,\si{\angstrom}$, reference energy $\epsilon_{\mathrm{ref}}/k_{\mathrm{B}}=1\,\si{\kelvin}$ and reference mass $m_{\mathrm{ref}}=1\,\si{\amu}$.  Consequently, the time reference
is $t_{\mathrm{ref}}=\sigma_{\mathrm{ref}} \sqrt{m_{\mathrm{ref}}/\epsilon_{\mathrm{ref}}}$.  The liquid phase was initially in its saturated state at the temperature $T=0.9$.  The vapor had a
temperature $T=0.8$, while its initial density varied between  $90\%$, $70\%$ and $50\%$ of the respective saturation density, cf. \cref{tbl:initialdata}.
\begin{table}[ht]
\caption{Initial data for the two-phase shock tube problem.}
\centering
\begin{tabular}{l|c|c|c|c||c|c|c|c}
\toprule
 & $\rho_{\liq}$ & $v_{\liq}$ & $T_{\liq}$ & $\lambda_{\liq}$ & $\rho_{\vap}$ & $v_{\vap}$ & $T_{\vap}$ & $\lambda_{\vap}$ \\
\midrule
Case 1 ($90\%$)   & 0.6635 & 0 & 0.9 & 4.2481 & 0.017800 & 0 & 0.8 & 0.31137   \\
Case 2 ($70\%$)   & 0.6635 & 0 & 0.9 & 4.2481 & 0.013844 & 0 & 0.8 & 0.31137   \\
Case 3 ($50\%$)   & 0.6635 & 0 & 0.9 & 4.2481 & 0.009889 & 0 & 0.8 & 0.31137  \\
\bottomrule
\end{tabular}
\label{tbl:initialdata}
\end{table}

The numerical models described in  Section \ref{sec:numericalmethods} were applied in CFD simulations: the sharp interface method, which solves the two-phase Riemann problem at the interface, and the
HEM. For the two-phase Riemann problem, the mass resistivity was specified as $\Rnn=200$ in all cases. This value was found by fitting the solution to the mass flux sampled by the present molecular
dynamics simulations, as no appropriate theoretical estimation of $\Rnn$ could be found. 

The computational domain was $x=(-400,1500)$ with the position of the interface at $x=0$.  The domain was discretized into $190$ grid elements with a piecewise polynomial approximation of degree
$N=3$, which results in fourth order accuracy in smooth parts of the flow. As boundary conditions, constant values were prescribed as Dirichlet condition.  Note that for the HEM simulation,
the solution was produced only by a second order FV method to enhance stability.  The Rusanov flux was used as numerical flux function and time integration was performed explicitly by a fourth order
Runge-Kutta scheme with five stages.  All simulations were performed until $t_{\mathrm{end}}=600$ was reached.

\begin{figure}[h!t]
   \centering
   \includegraphics[width=\linewidth]{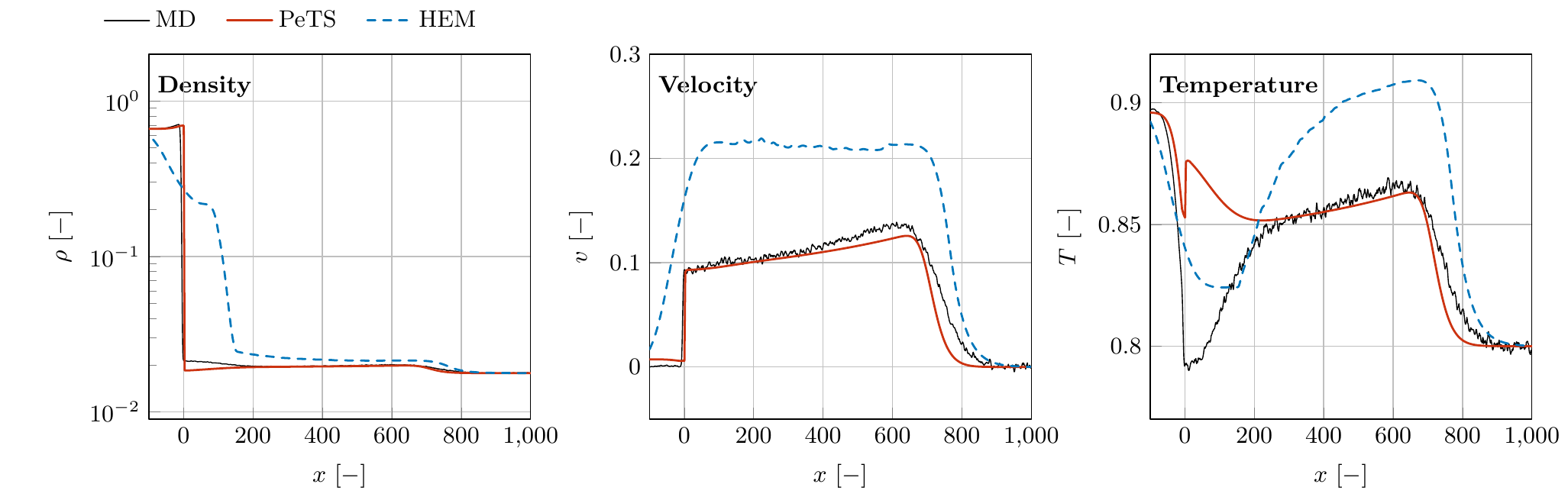}
   \caption{Results for case 1 with a vapor density of $90\%$ of its saturation value.
      The plot shows profiles of density, velocity and temperature at $t=600$ obtained from the sharp interface method, HEM and molecular dynamics.}
   \label{fig:results_case1}
\end{figure}

Results for case 1 with $90\%$ of the saturated vapor density are shown in \cref{fig:results_case1}.
Obviously, the HEM simulation is not able to reproduce the results of the molecular dynamics simulation.  The predicted mass flux is too large, which leads to wrong values for the
velocity and temperature in the vapor phase.  Furthermore, in the interfacial region, a rarefaction wave appears, which isentropically expands the liquid through the two-phase region towards a stable
vapor phase.  Over time, the extent of the rarefaction wave increases such that the interfacial region smears out over several orders of magnitude larger than evidenced by molecular dynamics.  While 
cooling effects in the liquid and vapor phases in the vicinity of the interface were observed, these processes were not driven by heat conduction, but are  rather a result of the isentropic expansion
of the wet vapor.

The sharp interface method shows a very good agreement with the molecular dynamics data in the homogeneous bulk phases, except for the vicinity of the interface.  The shock speed in the vapor phase as
well as the plateau values of density, velocity and temperature match with the molecular dynamics data. Due to the consideration of heat conduction in the bulk phases, the velocity gradient was
reproduced in the entire vapor phase.  On the liquid side, the very steep temperature gradient observed in the molecular dynamics data was also reproduced very well.  As the rarefaction wave already
moved out of the computational domain at $t=600$, this effect is a consequence of heat conduction.  This is also indicated by the increase of the liquid density towards the interface.

On the vapor side, the temperature profile predicted by molecular dynamics decreases with a large gradient towards the interface.  Consequently, an increasing density towards the interface is found.
The temperature on both sides of the interface is very similar. This fact validates the simplifications we made in the kinetic relation in \cref{eq:kineticrelation_isothermal_finished}.  In the
solution of the sharp interface method, the vapor temperature increases towards the interface, where a temperature jump between the adjacent liquid and vapor is observed. This jump does not violate
the continuity condition \eqref{eq:heat_coupling_2}, since it was only enforced in the subgrid model. For the actual solution of the Riemann problem, temperature jumps across the interface are
allowed. The temperature increase of the CFD solution towards the interface  occurs roughly in the vicinity of a change in the temperature gradient of the molecular dynamics data ($x\approx 200$).
Within this region, we suspect that local non-equilibrium effects take place which cannot be reproduced by our subgrid model relying on Fourier's heat conduction.  In addition, it is questionable
whether a macroscopic simulation can reproduce the molecular dynamics data in the vicinity of the interface at all. The basic assumption in the macroscopic solution is a discontinuous transition
between liquid and vapor. In some distance from the interface, local thermodynamic equilibrium is attained and the CFD solution that relies on  Fourier's heat conduction produces a very good
agreement with the molecular dynamics data.

\begin{figure}[h!t]
   \centering
   \includegraphics[width=\linewidth]{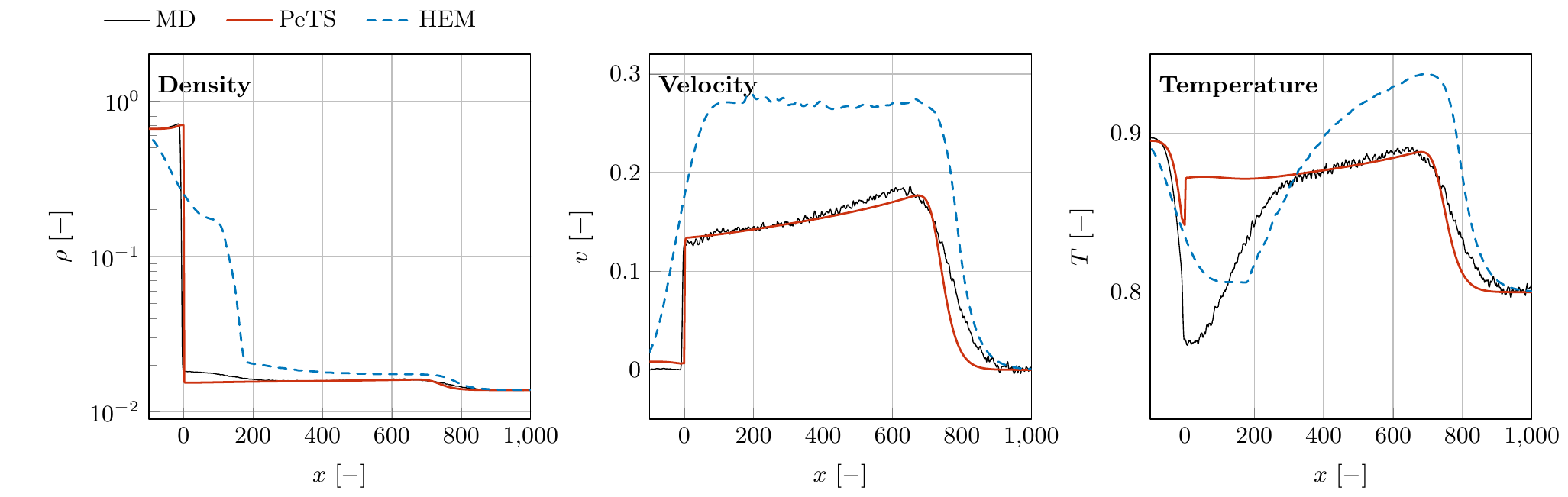}
   \caption{Results for case 2 with a vapor density of $70\%$ of its saturation value.
      The plot shows profiles of density, velocity and temperature at $t=600$ obtained from the sharp interface method, HEM and molecular dynamics.}
   \label{fig:results_case2}
\end{figure}

The results for case 2 with $70\%$ of the saturated vapor density are shown in \cref{fig:results_case2}.
Here, the mass flux is increased and hence density and velocity of the vapor phase are increased as well, while the propagation of the shock wave remains unaffected.  The rarefaction wave produced by
the HEM simulation increases its extent, especially in the vapor phase.  The sharp interface method compares well to the molecular dynamics data in the homogeneous phases, although the mass
resistivity $\Rnn$ was kept constant. The molecular dynamics data show that increasing the mass flux leads to a stronger cooling of the interfacial region. The same effect was observed with the sharp
interface method, where the temperature in this case remained nearly constant towards the interface.  The temperature jump between liquid and vapor remains, but decreases in its magnitude.

\begin{figure}[h!t]
   \centering
   \includegraphics[width=\linewidth]{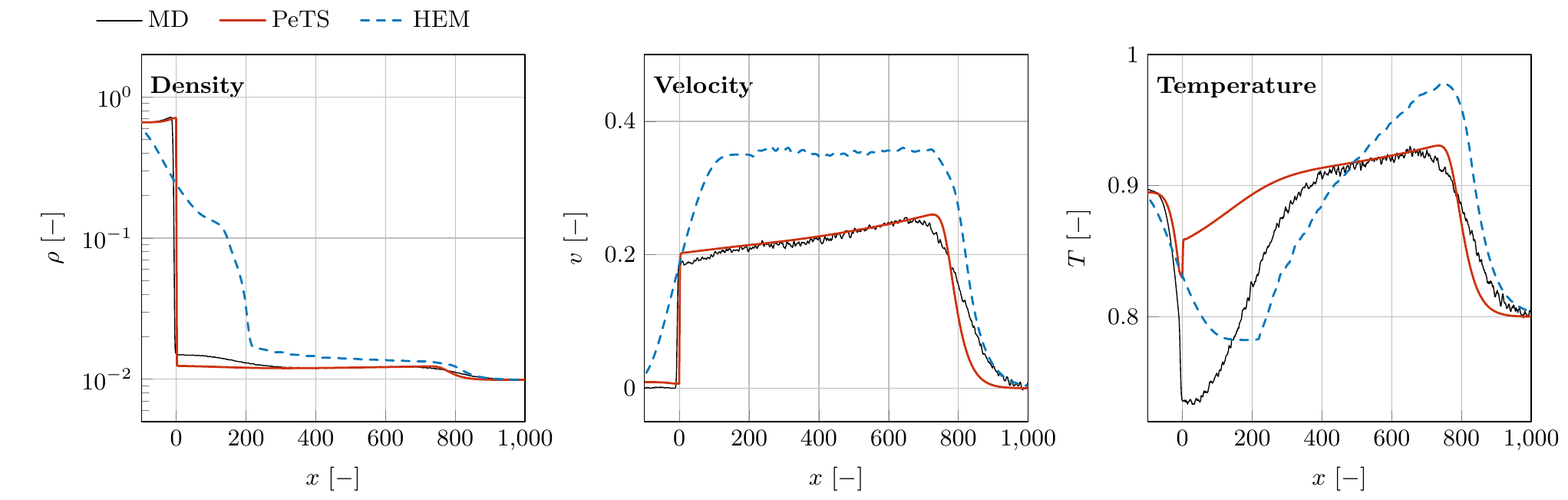}
   \caption{Results for case 3 with a vapor density of $50\%$ of its saturation value.
      The plot shows profiles of density, velocity and temperature at $t=600$ obtained from the sharp interface method, HEM and molecular dynamics.}
   \label{fig:results_case3}
\end{figure}

The results for case 3 with $50\%$ of the saturated vapor density are shown in \cref{fig:results_case3}.
The mass flux is increased further and the effects observed for case 2 are emphasized.  In the HEM profiles, the extent of the rarefaction wave increases even more, moving its impact on density and
temperature to the shock wave in the gas phase. For this more violent scenario, the sharp interface method still compares very well with the molecular dynamics data in the homogeneous phases, despite
using the same mass resistivity $\Rnn$ as in test cases 1 and 2.  In the interfacial region, cooling of the vapor phase becomes stronger. The sharp interface method now predicts a cooling effect as
well, but  fails to reproduce its correct extent. The temperature jump in the CFD data at the interface is less pronounced.

\section{Conclusion}
\label{sec:conclusion}
In this paper,  we continued to compare macroscopic simulations and solutions of the Riemann problem with molecular dynamics data for test cases, which were set up  such that non-equilibrium phase
transition occurs. The Lennard-Jones truncated and shifted model fluid was chosen for this task because it is computationally cheap in molecular dynamics, while highly accurate EOS are available for
the CFD side. The PeTS EOS of \citet{heier_equation_2018} was implemented in Fortran for this purpose and can be found in Ref. \cite{petsdata}.  We considered two types of macroscopic continuum
solutions: A sharp interface method combined with a level-set ghost fluid method based on a two-phase Riemann solver and the homogeneous equilibrium method, which heavily relies on equilibrium
assumptions. For the two-phase Riemann problem, a novel kinetic relation was proposed, considering basic ideas of non-equilibrium thermodynamics. Both methods were implemented into the DGSEM framework
\flexi\footnote{https://www.flexi-project.org/}. 

Comparing the results between the three methods revealed that  the HEM method is not capable to accurately reproduce the molecular dynamics data for the considered problems.  This is shown by the
overestimated mass flux and strong deviations for density, velocity and temperature. In addition, the appearance of a split rarefaction wave in the HEM is not evidenced by molecular dynamics. A
failure of the HEM method was expected because the phase equilibrium assumption does not hold in any of the considered cases. The sharp interface method, however, is able to reproduce many aspects of
the molecular dynamics data. First and foremost, the solutions indicate  that the  assumed structure of the Riemann solution is correct as the observed wave patterns in the molecular dynamics data and
from the sharp interface method coincide. In addition, the evaporation mass flux as well as the solution in the bulk  phases  share a good agreement with each other.  However, the temperature profile
in the vicinity of the interface deviates from the molecular dynamics data. The liquid phase is being cooled towards the interface, but this effect is weaker.  In addition, a temperature jump at the
interface can be observed. This entails that the vapor side has a significantly higher temperature than the liquid side. However, the molecular dynamics data show that both sides of the interface
exhibit a very similar temperature. 

As the mass flux and thus the macroscopic effect of the phase transition onto the bulk phases is replicated well, we assume that the differences in the temperature profile are not directly linked to
the proposed kinetic relation. It is more plausible that the subgrid model imposed for the heat flux closure needs further improvement.  In addition, it is possible that the application of Fourier's
law at the interface is incorrect and  a different heat conduction law needs to be used. Finally, we note that it is an open question whether macroscopic solutions are at all capable of
reproducing the temperature profile at the interface because strong non-equilibrium effects are at play here. 

To answer that question, we will continue the comparison of CFD with molecular dynamics considering stationary phase transitions. In comparison to the Riemann problem tackled in this work, 
convection would play a minor role and thermodynamic modelling could be thoroughly tested.  An approach to gain a consistent advection-diffusion solution of the two-phase Riemann problem based on Ref.
\cite{Joens2019} is currently under preparation. Also, comparisons with a fully hyperbolic flow model, i.e. the Godunov-Peshkov-Romenski equations \cite{Dumbser2015}, are planned to investigate the
applicability of Fourier's law.

\section*{Acknowledgements}
The work was supported by the German Research Foundation (DFG) through the Project SFB-TRR 75, Project number 84292822 - ``Droplet Dynamics under Extreme Ambient Conditions'' and under Germany's Excellence Strategy - EXC 2075 – 390740016. The simulations were performed on the national supercomputer
Cray XC40 (Hazel Hen) at the High Performance Computing Center Stuttgart (HLRS).




  \bibliographystyle{elsarticle-num-names} 
  \bibliography{two_phase_shocktube}


%
%
%
\end{document}

%% file: tikz_settings.tex
\usepackage{tikz}
\usepackage{xifthen}
\usetikzlibrary{spy,arrows}
\usetikzlibrary{arrows.meta}
\usetikzlibrary{shapes.geometric}
\usetikzlibrary{intersections}
\usetikzlibrary{positioning}
\usetikzlibrary{backgrounds}
\usepackage{pgfplots}
\pgfplotsset{compat=1.13}
\usetikzlibrary{calc}
\tikzstyle{rounded}  = [rounded corners]     
\tikzstyle{function} = [rectangle, text centered, draw=black,minimum width=6.0cm,minimum height=1.0cm]
\tikzstyle{mpi}      = [anchor=south west, xshift=0.2cm] 

\usetikzlibrary{3d}

\usetikzlibrary{fillbetween}
\usetikzlibrary{decorations,decorations.text,backgrounds}

\newcommand{\tikzAngleOfLine}{\tikz@AngleOfLine}
  \def\tikz@AngleOfLine(#1)(#2)#3{%
  \pgfmathanglebetweenpoints{%
    \pgfpointanchor{#1}{center}}{%
    \pgfpointanchor{#2}{center}}
  \pgfmathsetmacro{#3}{\pgfmathresult}%
}



%% file: commands.tex
\renewcommand{\l}{\left}
\renewcommand{\r}{\right}

\newcommand{\gradient}{\nabla}

\renewcommand{\div}{\nabla \cdot}

\newcommand{\transpose}{^{\top}}

\newcommand{\half}{\frac{1}{2}}
\renewcommand{\d}{\operatorname{d}}

\newcommand{\oneover}[1]{\ensuremath{\frac{1}{#1}}}

\renewcommand{\v}{\mathbf{v}}
\newcommand{\U}{\mathbf{U}}

\newcommand{\x}{\ensuremath{x}}

\newcommand{\Fb}{\mathbf{F}}

\newcommand{\Tinterface}{\ensuremath{T_{\Gamma}}}
\renewcommand{\v}{\ensuremath{v}}
\newcommand{\ninterface}{n}
\newcommand{\vninterface}{v}
\newcommand{\surfacetension}{\ensuremath{0}}
\newcommand{\q}{\ensuremath{q}}
\newcommand{\qninterface}{\ensuremath{q}}

\newcommand{\Rnn}{\ensuremath{R_{\dot{m}}}}
\newcommand{\Rqq}{\ensuremath{R_{e}}}

\newcommand{\R}{\mathbb{R}}

\newcommand{\Rgasconstant}{\ensuremath{\mathcal{R}}}

\newcommand{\liq}{\ensuremath{\mathrm{liq}}}
\newcommand{\vap}{\ensuremath{\mathrm{vap}}}
\newcommand{\sat}{\ensuremath{\mathrm{sat}}}

\newcommand{\flexi}{\textit{FLEXI}\xspace}

\newcommand{\etal}{et al.~}